\documentclass[twocolumn,showpacs,preprintnumbers,amsmath,amssymb,prl]{revtex4}

\usepackage{graphicx}
\usepackage{dcolumn}
\usepackage{bm}
\usepackage[]{epsfig}

\begin{document}
\title{Origin of the slow dynamics and the aging of a soft glass}
\author{Sylvain Mazoyer, Luca Cipelletti and Laurence Ramos$^*$}
\affiliation{Laboratoire des Collo\"{\i}des, Verres et
Nanomat\'{e}riaux (UMR CNRS-UM2 5587), CC26, Universit\'{e}
Montpellier 2, 34095 Montpellier Cedex 5, France}

\email{ramos@lcvn.univ-montp2.fr}
\date{\today}

\begin{abstract}
We study by light microscopy a soft glass consisting of a compact
arrangement of polydisperse multilamellar vesicles. We show that
its slow and non-stationary dynamics results from the unavoidable
small fluctuations of temperature, which induce intermittent local
shear deformations in the sample, because of thermal expansion and
contraction. Temperature-induced shear provokes both reversible
and irreversible rearrangements whose amplitude decreases with
time, leading to an exponential slowing down of the dynamics with
sample age.
\end{abstract}

\maketitle


Quite generally, correlation functions in glassy systems exhibit a
two-step relaxation \cite{Donth2001}. This applies to supercooled
molecular liquids and spin glasses, but also to a large variety of
soft materials, such as concentrated colloidal suspensions
\cite{vanMegenPRE1998}, emulsions \cite{HugangPRE99}, surfactant
phases, and gels \cite{Faraday03}. In soft glasses, the initial
decay of density-density correlators measured e.g. by dynamic light
scattering is well understood: it corresponds to the thermally
activated motion of particles in the cage formed by their neighbors
\cite{vanMegenPRE1998}, or to overdamped phonons, whose amplitude is
restricted by structural constraints, like in concentrated emulsions
\cite{HugangPRE99} or surfactant phases \cite{Faraday03}. By
contrast, the motion associated with the final relaxation of the
correlation function is still poorly understood, in spite of the
large research effort of the last years. Subdiffusive
\cite{vanMegenPRE1998,SimeonovaPRL2004}, diffusive
\cite{AbouPRE2001}, hyperdiffusive \cite{KnaebelEPL2000} and
ballistic \cite{Faraday03,BandyopadhyayPRL2004} behavior has been
observed, often associated with dynamical heterogeneity
\cite{KegelScience2000,WeeksScience2000,LucaJPCM2003,DuriPRE2005}
and aging
\cite{KnaebelEPL2000,AbouPRE2001,ViasnoffPRL2002,CourtlandJPCM2003,
SimeonovaPRL2004}. In most cases, the origin of this slow dynamics
is not clear, although elasticity and the relaxation of internal
stress have been highlighted as possible key ingredients
\cite{Faraday03,RamosPRL2005,BandyopadhyayPRL2004,KnaebelEPL2000}.
Indeed, light scattering experiments coupled to rheology have shown
that an external oscillatory shear strain can help the system evolve
towards a more relaxed configuration, presumably by relaxing
internal stress
\cite{PineHebraudPRL,PetekidisPRE2002,ViasnoffPRL2002}. However, the
physical mechanism by which internal stress may be relaxed in
undriven systems has not been identified clearly to date.

In this Letter we present light microscopy experiments that probe
the aging dynamics of a compact arrangement of multilamellar
vesicles. Surprisingly, we find that the driving mechanism for the
slow evolution of the sample configuration are the experimentally
unavoidable fluctuations of temperature, which induce intermittent
local mechanical shears in the sample, due to thermal expansion
and contraction. We find that the amplitude of both reversible and
irreversible rearrangements provoked by the shear decreases with
time, leading to aging.

The sample is a water-based mixture of surfactants and a
block-copolymer, whose composition is given in
ref.~\cite{RamosPRL2001}. At $5^{\circ} \rm{C}$ the sample is
fluid; above $8^{\circ} \rm{C}$ the increase of the hydrophobicity
of one segment of the block-copolymer leads to the formation of
polydisperse multilamellar vesicles, or
onions~\cite{equilibriumonions}. All experiments are performed at
temperature $T=23.4^{\circ} \rm{C}$, where the onions are densely
packed and no changes in the structure are observed with time.
Previous linear rheology and dynamic light scattering experiments
\cite{RamosPRL2001,RamosPRL2005} have shown that this system
displays a slow dynamics whose characteristic timescale increases
with time after a transition from the fluid to the gel-like phase
(``aging'').

\begin{figure}
\includegraphics{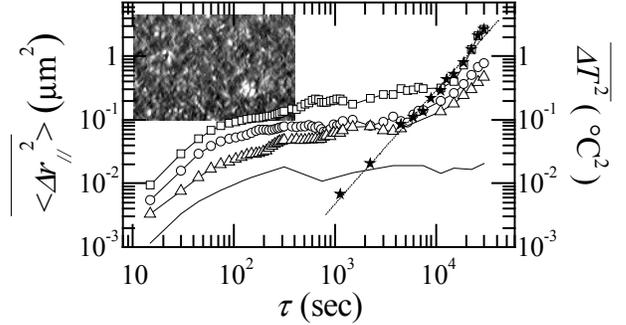}
\caption{Relative MSD along the long axis of the capillary as a
function of the time lag $\tau$. The sample age is $t_{w}=7500 \,
\rm{sec}$ (squares), $22500 \, \rm{sec}$ (circles) and $37500 \,
\rm{sec}$ (triangles). Stars: MSD calculated by taking into
account only the irreversible rearrangements. The dotted line is a
power law fit yielding an exponent $1.8\pm 0.1$. Full line: ``mean
squared displacement'' of the temperature fluctuations,
$\overline{\Delta T^2}$, as defined in the text.
Inset: Portion of a typical 
image 
of size $283 \, \mu \rm{m} \times 184 \, \mu \rm{m}$.}
\label{FIG:1}
\end{figure}

The sample is loaded at $5^{\circ} \rm{C}$ in a glass capillary of
length a few cm and rectangular cross-section ($0.2  \times 2 \,
\rm{mm}^2$), which is flame-sealed to prevent evaporation.
Centrifugation is used to confine the air bubble left after
filling at one end of the capillary. The sample is then placed in
an oven (Instec) that sets the temperature to $23.4^{\circ}
\rm{C}$ ($T$ is measured on the capillary in the close vicinity of
the sample). The standard deviation of $T$ over 1 day is
$0.09^{\circ} \rm{C}$. Age $t_{w}=0$ is defined as the time at
which $T$ has reached $23.4^{\circ} \rm{C}$. A microscope equipped
with a 10x objective is used to image the onions between crossed
polarizers (see the inset of Fig.~\ref{FIG:1}). The field of view
is $0.93 \, \rm{mm} \times 1.24 \, \rm{mm}$ and is located in a
horizontal plane in the center of the capillary. Images are taken
every $15$ sec for about $24$ hours and analyzed with the Image
Correlation Velocimetry method~\cite{PIV} to quantify the time
evolution of the (coarse-grained) displacement field. Each image
is divided into $192$ regions of interest (ROIs) corresponding to
$76 \, \mu \rm{m} \times 76 \, \mu \rm{m}$ in the sample and the
displacement (i. e. translational motion) $\Delta
\textbf{R}_i(t_{w},\tau)$ of each ROI for pairs of images taken at
time $t_{\rm w}$ and $t_{\rm w}+\tau$ is measured with an accuracy
of 50 nm \cite{NotePIV}.

\begin{figure}
\includegraphics{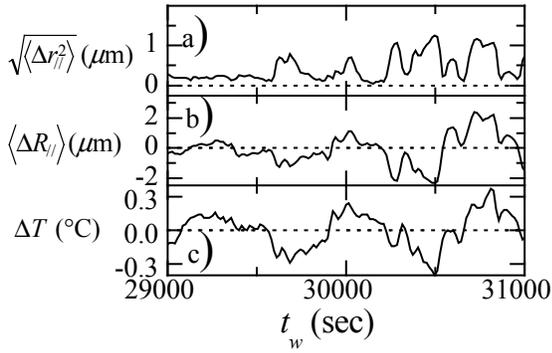}
\caption{Age dependence of a) the square root relative parallel
displacement, $ \sqrt{\langle \Delta r_{\parallel}(t_{\rm
w},\tau)^2 \rangle} $; b) the global parallel displacement
$\langle \Delta R_{\parallel}(t_{\rm w},\tau) \rangle $; c) the
temperature variation over a lag $\tau$, $\Delta T(t_{\rm
w},\tau)$. Note that $\Delta T$ shown here is among the largest
recorded during the experiment. In all panels $\tau = 315$ sec. }
\label{FIG:2}
\end{figure}

We first discuss the relative displacement, defined as $\Delta
\textbf{r}_i (t_{w}, \tau)=\Delta \textbf{R}_i (t_{\rm w},\tau) -
\langle \Delta \textbf{R}_i (t_{\rm w},\tau)\rangle$, where $\langle
\cdot\cdot\cdot \rangle $ denotes an instantaneous spatial average
over all ROIs. We calculate the mean-squared relative displacements
(MSDs) $\overline{\langle \Delta r_{\parallel}
(t_{w},\tau)^2\rangle}$ and $ \overline{\langle \Delta r_{\perp}
(t_{w},\tau)^2\rangle }$, where the subscript $i$ has been dropped
for simplicity and $\parallel$ and $\perp$ refer to the two
horizontal components of $\Delta \textbf{r}_i$, parallel and
perpendicular to the long axis of the capillary, respectively.
$\overline{\cdot\cdot\cdot}$ denotes a time average over a window of
$15000$ sec centered around $t_{\rm w}$. We find that the motion is
strongly anisotropic, with the parallel displacement typically one
order of magnitude larger than the perpendicular one; therefore, in
the following we focus on the parallel motion that dominates the
dynamics. Figure ~\ref{FIG:1} shows the parallel MSD as a function
of the lag $\tau$, for three ages (open symbols). Three distinct
regimes are observed: a first rapid increase of the MSD with $\tau$
at short lags, an almost flat plateau at intermediate lags,
extending over about two orders of magnitude in $\tau$, and a
long-lag regime where the MSD increases strongly with $\tau$.
Moreover, the dynamics slows down with age, as revealed by the
decrease of the MSD with $t_{\rm w}$ at all lags.

\begin{figure}
\psfig{file=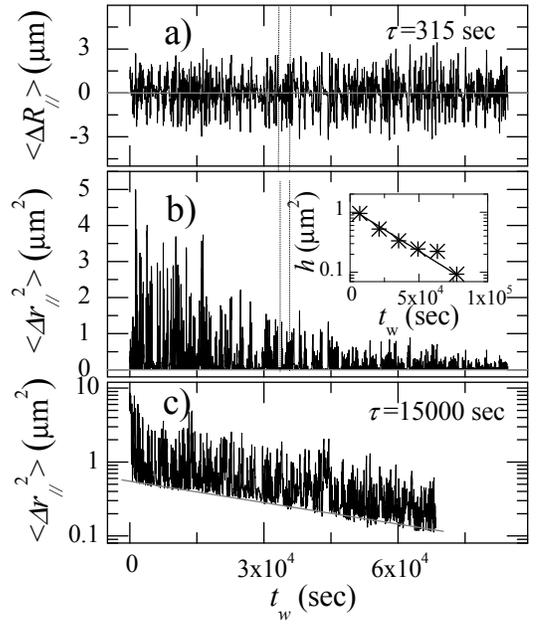,width=7.cm}
\caption{Age dependence of a) the spatially-averaged absolute
displacement at a lag $\tau = 315$ sec; b) and c) the relative
squared displacement at $\tau = 315$ and $15000$ sec,
respectively. The vertical dotted lines in a) and b) show the time
window expanded in figs.~\ref{FIG:2} a) and b). The finite base
line in c) (grey line) corresponds to irreversible rearrangements
that persist even when $\Delta T =0$. Inset of b): stars, mean
value of the top 25\% intermittent peaks of $\langle \Delta
r_{\parallel}^{2} \rangle$; line, exponential fit.}
 \label{FIG:3}
\end{figure}

To get insight in the physical mechanisms at the origin of this slow
dynamics, we examine time-resolved quantities. The two-time squared
relative displacement, $\langle \Delta r_{\parallel} (t_{\rm
w},\tau)^2\rangle$, exhibits strong fluctuations with $t_{\rm w}$,
with intermittent peaks of high amplitude. This is shown, for $\tau
= 315$ sec, in fig.~\ref{FIG:2}a for a short time window and in
fig.~\ref{FIG:3}b for the whole duration of the experiment (this lag
corresponds to the plateau of the MSD, see fig. 1). A visual
inspection of the movie of the sample motion suggests that the
fluctuations of the relative displacement are associated with those
of the global displacement, $\langle \Delta R_{\parallel}
(t_{w},\tau) \rangle $, and that the latter are of the same order of
magnitude as the former. The movie shows also a large drift during
the initial temperature jump, due to the thermal expansion of the
sample. This suggests that the fluctuations of $\langle \Delta
R_{\parallel} \rangle $ recorded once a constant temperature is
reached may be due to small fluctuations of $T$, which are
experimentally unavoidable. Indeed, a raise (decrease) of $T$ would
induce an expansion (contraction) of the sample. Because the
material is confined, it would deform essentially perpendicular to
the free interface. Hence, the expansion/contraction of the sample
is expected to be uniaxial, resulting mostly in a global
displacement in the $\parallel$ direction, as observed
experimentally.

\begin{figure}
\psfig{file=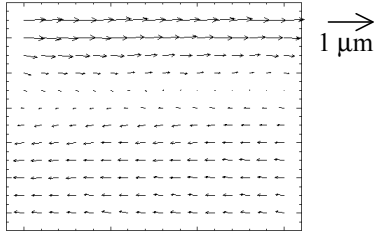,width=5.cm}
\caption{Relative displacement field between two images separated by
$315\, \rm{sec}$, for $t_{\rm w} = 2280 \, \rm{sec}$. The arrows
indicate the displacement for each ROI. The size of the images is
$0.93 \, \rm{mm} \times 1.24 \, \rm{mm}$, with the longest side
along the $\parallel$ direction.} \label{FIG:4}
\end{figure}

To test this hypothesis, we show in fig.~\ref{FIG:2}c the
variation of $T$ over a lag $\tau$, defined as $\Delta
T(t_{w},\tau)= T(t_{w}+\tau) - T(t_{w})$, and we compare it to $
\sqrt{\langle \Delta r_{\parallel}^2 (t_{w},\tau) \rangle} $
(fig.~\ref{FIG:2}a) and $\langle \Delta R_{\parallel} (t_{w},\tau)
\rangle $ (fig.~\ref{FIG:2}b). Clearly, whenever $\Delta T$
increases or decreases so does $\langle \Delta R_{\parallel}
\rangle $. Similarly, any peak in the relative mean squared
displacement (positive by definition) corresponds to a (positive
or negative) fluctuation of $\Delta T$. Note that the
time-averaged value of the global displacement $\langle \Delta
R_{\parallel} \rangle$ is $0$ (see fig.~\ref{FIG:3}a),
consistently with the fact that there is no net drift of $T$. The
similarity between the different signals shown in fig.~\ref{FIG:2}
can be quantified by calculating their linear correlation
coefficient, $c$, which ranges from 0 to 1. For the pair $\Delta
T(t_{w})$, $\langle \Delta R_{\parallel}(t_{w}) \rangle$ and for
all ages and all lags, we find $c = 0.74 \pm 0.15$ demonstrating a
strong correlation between the fluctuations of the global
displacement field and those of the temperature. A significant
correlation, although with a lower value $c = 0.57 \pm 0.15$, is
also found between $\sqrt{\Delta T^2}$ and $\sqrt{\langle \Delta
r_{\parallel}^2 \rangle}$, suggesting that the larger the global
displacement field, the more spatially heterogeneous the
deformation. This heterogeneity is in contrast to what may be
expected for simple fluids or diluted suspensions. It is likely to
stem from local variations of the viscoelastic properties of the
sample, due to its jammed nature, and to the curvature of the
meniscus at the sample-air interface. For lags corresponding to
the MSD plateau, the relative displacement field most often
corresponds to a shear in the $\parallel$ direction, as shown in
fig.~\ref{FIG:4}. Note that the displacement field varies slowly:
thus, the dynamics is strongly correlated in space and the shear,
of order $0.1 \%$ at most, is in the linear regime, as determined
by a strain sweep at a frequency of $1$ Hz ~\cite{RamosPRL2001}.
This long-range correlation is in contrast with the case of
molecular glass formers and colloidal hard spheres, for which
dynamical correlations extend over a few
particles~\cite{Donth2001,WeeksScience2000}.

To seek further support for the crucial role of the fluctuations
of $T$, we compare the macroscopic thermal expansion coefficient,
$\chi_{T}$, to that estimated from our observations. For the
latter, one has $\chi_{T}=\frac{1}{L}\frac{\delta L}{\delta T}$,
where $L = 2 $~cm is the length of the sample from the bubble-free
capillary end to the position of the field of view and $\delta L$
the length variation provoked by a temperature variation $\delta
T$. By taking for $\delta T$ and $\delta L$ the standard deviation
of $\Delta T (t_w,\tau)$ and $\langle\Delta R_{\parallel}
(t_w,\tau) \rangle$, respectively, and using the data for $\tau =
315$ sec we find $\chi_{T}= 40 \pm 10 \, \rm{mK}^{-1}$ in good
agreement with $26 \, \rm{mK}^{-1}$ for water
\cite{ThermalExpansionCoeff}.

Temperature fluctuations are thus at the origin of the
intermittent motion shown in figs.~\ref{FIG:3}a,b, where peaks of
$\Delta T$ correspond to large fluctuations of $\langle\Delta
R_{\parallel} (t_w,\tau) \rangle$ and hence of $ \langle \Delta
r^{2}_{\parallel} \rangle$. Given the key role of temperature
fluctuations, deeper insight on the evolution of the MSD shown in
fig.~1 may be obtained by comparing the $\overline{\langle \Delta
r^{2}_{\parallel} \rangle}$ data to the analogous of the MSD for
$T$, $ f_T(\tau)= \overline{\Delta T(t_{w}, \tau)^2} $. This
quantity is plotted in fig.~\ref{FIG:1} as a solid line and
displays remarkable analogies with the MSD. Similarly to the MSD,
$f_T(\tau)$ increases with $\tau$ until reaching a plateau for
$\tau = \tau_{c} \sim 300$ sec, a time scale of the order of the
slowest fluctuations of $T$. For $\tau < \tau_{c}$, $T$ is on
average monotonic. The relative MSD follows the evolution of $T$
and hence increases monotonically as well. By contrast, for time
lags $\tau \gtrsim \tau_{c}$, the sample has been submitted to
several fluctuations of $T$, which have induced several shear
deformations in the two opposite directions. On these time scales,
this back and forth motion is virtually fully reversible and does
not lead to a growth of the cumulated displacement; hence, a
plateau is measured for the MSD. The reversible nature of the
motion in this regime is further supported by the observation that
the peaks of the two-time squared displacement shown in
fig.~\ref{FIG:3}b raise from a base line that is essentially zero,
corresponding to pairs of images taken
---by chance--- at nearly the same $T$. Thus, the initial
growth of $\overline{\langle \Delta r^{2}_{\parallel} \rangle}$ and
the plateau are not due to the usual ``rattling in the cage''
mechanism, but rather to reversible shear deformations caused by
temperature fluctuations. However, while $f_T(\tau)$ saturates at
the plateau value for all $\tau > \tau_{c}$, a further increase of
the MSD is observed at very large $\tau$. This suggests that several
expansion/contraction cycles imposed by the fluctuations of $T$
eventually trigger irreversible rearrangements, whose cumulated
effect is responsible for the final growth of the MSD.

This picture is confirmed by the time evolution of $ \langle
\Delta r_{\parallel}^2 (t_{w},\tau) \rangle $ calculated for a lag
$\tau=15000 \, \rm{sec} \gg \tau_{c}$ and shown in
fig.~\ref{FIG:3}c. Intermittent peaks similar to those of
fig.~\ref{FIG:3}b are observed, corresponding to the largest
values of $\Delta T(t_{w},\tau)$. In contrast with the data for
shorter lags, however, the peaks in fig.~\ref{FIG:3}c raise from a
base line larger than zero (grey line). As discussed previously,
the baseline corresponds to pairs of images taken at nearly the
same $T$, in the absence of any global displacements. A non-zero
baseline confirms that the $T$-induced shear deformations have led
to irreversible rearrangements of the sample. This behavior is
strongly reminiscent of that of colloidal systems submitted to a
mechanical shear
\cite{PineNature,PineHebraudPRL,PetekidisPRE2002,ViasnoffPRL2002,OzonPRE2003}
or vibrated granular media \cite{granular}. Although the
microscopic mechanism responsible for rearrangements may vary
(e.g. hydrodynamic interactions play a major role in
\cite{PineNature,PetekidisPRE2002}), the general picture is the
same: shear deformations lead to irreversible rearrangements. In
our experiments no mechanical shear is imposed; remarkably,
however, the small strain $\sim 10^{-3}$ due to temperature
fluctuations $\delta T \sim 0.1^{\circ} \rm{C}$ is sufficient to
induce irreversible rearrangements on very long time scales.

At a given age, the height of the baseline shown in fig. 3c
provides a direct measurement of the relative mean squared
displacement for $\tau = 15000$ sec when only irreversible
rearrangements are taken into account. We calculate the
irreversible mean squared displacement, $\rm{MSD}_{\rm
irr}(\tau)$, by analyzing in a similar way data at a variety of
time lags, and show in fig.~\ref{FIG:1} the result for $t_{\rm w}
= 7500$ sec (star symbols). Remarkably, we find $\rm{MSD}_{\rm
irr} \sim \tau^p$ over about two orders of magnitude in $\tau$,
with $p=1.8\pm 0.1$, thus indicating that the irreversible motion
is close to ballistic, for which $p=2$. We stress that a similar
ballistic motion has been invoked to explain the slow dynamics in
a variety of soft glassy materials probed by light and X-photon
scattering techniques
\cite{Faraday03,RamosPRL2001,BandyopadhyayPRL2004}, including the
very same samples described here.

With age, the change in configuration corresponding to both the
reversible and the irreversible rearrangements diminishes: this is
demonstrated by the decrease of the height of the peaks of $
\langle \Delta r_{\parallel}^2\rangle $ with $t_{\rm w}$
(figs.~\ref{FIG:3}b,c), as well as by the negative slope of the
base line of $ \langle \Delta r_{\parallel}^2\rangle $ measured at
long lags (fig.~\ref{FIG:3}c). We find that the amplitude of both
reversible and irreversible rearrangements decays exponentially
with sample age, with roughly the same characteristic time
$\tau_{\rm aging}$: independently of $\tau$, $\tau_{\rm aging}
\sim 39900 \pm 5500 \, (30200 \pm 4300)$ sec for irreversible
(reversible) rearrangements. For irreversible rearrangements,
$\tau_{\rm aging}$ is obtained from the slope of the base line in
a log-lin representation of $ \langle \Delta
r_{\parallel}^2\rangle $ \textit{vs} $t_{\rm w}$, as in fig. 3c.
For reversible rearrangements, we divide the data in 6 time
windows and plot the average height, $h$, of the top 25\% largest
peaks of each window as a function of $t_w$ (see inset of fig.
3b); $\tau_{\rm aging}$ is then obtained by fitting an exponential
decay to $h(t_{\rm w})$~\cite{NotePercentage}. A similar
exponential aging has been observed for other soft glasses
\cite{Faraday03,AbouPRE2001}; it may be explained by assuming
$\rm{MSD}_{\rm irr} \propto n_{\rm rearr}$, where $n_{\rm rearr}$
is the number of irreversible rearrangements per unit time,
occurring in ``weak sites''. These sites correspond to the regions
that are initially most unstable mechanically and are more likely
to be relaxed by a rearrangement. We further assume $n_{\rm rearr}
= \Gamma N(t)$, with $\Gamma$ the probability of rearrangement per
unit time (dictated by the fluctuations of temperature and thus
independent of age) and $N(t)$ the number of weak sites not yet
rearranged. The temporal evolution of $N$ obeys $dN(t)/dt =
-\Gamma N(t)$ and hence $N(t) \propto \exp (-\Gamma t)$, thus
explaining the exponential decay of the irreversible mean square
displacement. We cannot exclude that new weak sites are created,
however, in the time window we probe, the dynamics is dominated by
the relaxation process. Note that the amplitude of the deformation
field due to a temperature-induced expansion or contraction is
likely to be larger the more numerous the weak zones still to be
relaxed. Thus, the evolution of $N(t)$ would also explain the
exponential decay of $h$ that parallels that of $\rm{MSD}_{\rm
irr}$.

As a final remark, we note that the behavior reported in this work
should be relevant to other closely packed soft systems, such as
concentrated emulsions or surfactant phases and swollen polymer
spheres or star polymers, where elasticity prevails over
dissipation. Indeed, $T$ fluctuations in most typical experiments
on soft materials are comparable to those reported here and would
lead to similar strains, since the thermal expansion coefficient
of water- or organic solvent-based systems is of the same order of
magnitude.

We thank W. Kob, L. Berthier, and E. Pitard for discussions. This
work was supported in part by the European MCRTN ``Arrested matter''
(MRTN-CT-2003-504712) and NoE ``SoftComp`` (NMP3-CT-2004-502235),
and by CNES (03/CNES/4800000123) and the Minist\`{e}re de la
Recherche (ACI JC2076). L.C. is a junior member of the Institut
Universitaire de France.



\begin{references}
\bibitem{NotePIV} This technique does not allow one to directly
probe internal motions on a length scale smaller than the ROI
size. By changing the ROI size, we have checked that the internal
motion within ROIs is negligeable compared to the translational
motions of the ROIs.
\bibitem{ThermalExpansionCoeff} Recall that the onion phase is
composed by $84 $\% water.
\bibitem{NotePercentage} Similar results are obtained with other threshold values.
\end{references}
\end{document}